\documentclass{ws-ijmpb}

\begin{document}

\markboth{Chia-Ren Hu}
{Qualitative Picture of a New Mechanism for High-$T_c$
Superconductors}

\catchline{}{}{}{}{}
%
%%%%%%%%%%%%%%%%%%%%%%%%%%%%%%%%%%%%%%%%%%%%%%%%%%%%%%%%%%%%%%%%%%%%

\title{QUALITATIVE PICTURE OF A NEW MECHANISM FOR \\
HIGH-$T_c$ SUPERCONDUCTORS}

\author{CHIA-REN HU}

\address{Department of Physics, Texas A\&M University,
\\College Station, TX 77843-4242, U.S.A.\\
crhu@tamu.edu}

%\author{SECOND AUTHOR}
%
%\address{Group, Laboratory, Address\\
%City, State ZIP/Zone, Country\\
%second\_author@group.com}

\maketitle

\begin{history}
\received{(16 January 2003)}
\revised{(31 January 2003)}
%\accepted{(Day Month Year)}
%\comby{(xxxxxxxxxx)}
\end{history}

\begin{abstract}
Xu et al. observed enhanced Nernst effect and Iguchi et al.
observed patched diamagnetism, both well above $T_c$ in
underdoped high-$T_c$ superconductors (HTSCs). A new mechanism
is proposed here, which seems to naturally explain, at least
qualitatively, these observations, as well as the d-wave nature
and continuity of pseudogap and pairing gap, the tunneling
conductance above $T_c$, as well as $T^*(x)$, $T_{\nu}(x)$,
$T_c(x)$, etc. This mechanism combines features of dynamic
charged stripes, preformed pairs, and spin-bags: At appropriete
doping levels, the doped holes (and perhaps also electrons) will
promote the formation of anti-phase islands in short-range
anti-ferromagnetic order. On the boundary of each such island
reside two doped carriers; the unscreened Coulomb repulsion
between them stabilizes its size. Superconductivity results
when such ``pre-formed pairs'' Bose-condense.
\end{abstract}

\vspace*{12pt}

%\section{General Appearance}    %) A SECTION HEADING

Recently Xu et al.\cite{Xu-etal} observed enhanced Nernst effect
well above $T_c$ in underdoped high-$T_c$ cuprates. It occurs
below an onset temperature $T_{\nu}$, which first rises sharply
at very low hole concentrations ($x$), reaching a peak  well
below optimal doping, and then decreases monotonically as $x$
increases further. (For La$_{2-x}$Sr$_x$CuO$_4$, or LSCO, the
peak $T_{\nu}$ is $\sim$ 128 K whereas the maximum transition
temperature $T_c$ is below 40 K.) The Nernst effect is
measured as follows: A rectangular slab of a single-crystal
sample has its long edges parallel to the $x$ and $y$ axes. A
magnetic field $B_z$ is applied along $z$, and a temperature
gradient
$\partial_x T$ is applied along $x$. The Nernst signal is the
$B_z$-antisymmetric electric field $E_y$ detected along $y$.
The Nernst coefficient is defined as
\begin{equation}
\nu = \frac{1}{B_z}\frac{E_y}{\partial_x T}\,.
\label{Nernst}
\end{equation}
It is well-known that enhanced Nernst effect can be observed
around and below $T_c$ of a low-$T_c$, type-II
superconductor.~\cite{low-Tc} It has been explained in terms
of the vortex lines in the superconductor. The core of each
vortex line has low-lying quasiparticle states much below the
sueprconducting gap.  Thus there is extra entropy localized
inside the vortex core. A positive $\partial_x T$ then makes
the vortex lines move toward $-x$, because an entropy current
times $T$ is a heat current, and heat always flows from hot to
cold. But a vortex line is also a flux tube, Thus as the vortex
lines move toward $-x$ there is flux cutting through any line
along $y$, giving rise to a positive electric field along $-y$.
This ``flux-flow'' origin of the enhanced Nernst effect in
low-$T_c$ superconductors near and below $T_c$ has been
established since more than thirty years ago. Thus when Xu et
al. observed enhanced Nernst effect in HTSCs, they naturally
associated it with vortices also, even though this time the
effect was observed well above $T_c$. (But they did carefully
say ``vortex or {\it vortex-like} excitations.~\cite{PALee}.)
Another recent experimental work, however, appears to have given
direct evidence that this enhanced Nernst effect is not due to
vortices. Iguchi et al.\cite{Iguchi-etal}, using scanning SQUID
microscopy, observed vortices below $T_c$ only, and patched
diamagnetism well above $T_c$ ($= 18 - 19$ K) up to as high as
80K in LSCO with $x \simeq 0.10$ (for which $T_{\nu}$ is above
120 K.). They also briefly stated that for a nearly
optimally-doped sample, similar patched diamagnetism is also
observed for about 5 K above $T_c$ ($\simeq$ 40 K.) Even in a
slightly under-doped YBCO thin-film sample ($T_c \simeq 84 K$),
similar patched diamagnetism is also observed for a limited
range of $T$ above $T_c$. The distinction is very unambigious.
I therefore feel that these two observations taken together are
not explained by any of the high $T_c$ mechanisms proposed to
date. Thus I venture to propose a new mechanism here (only a
qualitative one so far).

I interprete Ref.~\cite{Iguchi-etal} as to mean that the
system contains {\it localized} objects which are diamagnetic and
are capable of phase separation. $T_{\nu}$ is the formation
temperature of these objects (as in the the formation of
molecules from the constituent atoms), and the number of these
objects should increase as the temperature is lowered from
$T_{\nu}$, in order to account for the observed growing total
area of these diamagnetic patches. Reference~\cite{Xu-etal} then
implies that these objects are also associated with local entropy.
In order for these objects to give an enhanced Nernst effect with
the same sign as that generated by vortices, these objects must
be associated with local {\it depletion} of entropy rather than
local excess of entropy. This is because a vortex is associated
with localized flux along $B_z$, whereas a diamagnetic object has
induced flux opposite to $B_z$. Thus our main aim is to find a
mechanism by which such objects can be generated.

As is already widely accepted, the antiferromagnetism observed
in the cuprates at zero and very low doping originates from a
Mott-Hubbard insulating state. It can be described by a
single-band Hubbard model.~\cite{anderson} At such very low
doping the very low concentration of doped holes --- perhaps
also electrons; the difference has not yet been examined --- are
likely to form {\it localized, immobile,} singly-charged magnetic
polarons (also known as spin-bags).~\cite{spinbag} Such localized
immobile objects clearly can not destroy long-range
antiferromagnetic order (AFO). The lack of mobility of these
charged objects at very low concentrations is most likely due to
the inhomogeneous potential generated by the dopant ions, which
keeps the well-separated charge carriers trapped to the vicinity
of their parent dopant ions. As doping increases to the extend
that the mean carrier separation becomes comparable to the
distance between the dopant ions and the nearest CuO$_2$ planes,
the inhomogeneity of the potential becomes very week. The doped
carriers should then become mobile at $T = 0$. This should be
roughly when antiferromagnetism disappears at $T = 0$, since it
is well known that a moving hole leaves a trail of ``wrong
spins'' with respect to the AFO. Indeed long-range AFO
disappears in these materials at a very low hole concentration
$x = x_{c1} = 0.02 - 0.03$ (and a bit larger electron
concentration). Thermal energy can help carriers to become
mobile at lower $x$ than $x_{c1}$, and therefore long-range AFO
disappear at increasingly smaller $x$ as $T$ is increased.

As soon as some doped charge carriers become mobile, they can
reorganize to form new objects. I propose that these new objects
are doubly-charged anti-phase islands (DCHAPHIs) in a
short-range AFO background. The boundary of such an island is
like a dynamic stripe, which is known to be able to trap
charges,~\cite{stripe} but now bent around to form a small
closed loop rather than extending from one edge of the sample to
the opposite edge. The ``loop'' is so small that only two holes
(and perhaps also electrons) can be trapped on it, in a
spin-singlet state, so that the two charges can occupy the same
lowest-energy orbital state on the loop. Any bound pairs formed
above the superconducting transition temperature have been
called ``preformed pairs''. But the ``preformed pair'' proposed
here are not bound to each other by any attractive force between
them, but are rather both trapped on the same boundary of a
single anti-phase island. (In this sense they resemble a
doubly-charged spin bag.~\cite{spinbag}, except that the orbital
state in the present case is likely to be $d$-wave (see below),
whereas the orbital state in a spin bag is likely to be
$s$-wave,~\cite{spinbag} since the bag is presumably deepest at
its center. Coulomb repulsion between the two charges may
convert it to $d$-wave, but once both charge carriers avoid the
center region in order to avoid each other, I expect AFO to
recover in that region to turn a doubly-charged spin-bag into
a DCHAPHI.) The DCHAPHIs are expected to be also associated with
local depletions of {\it magnetic} entropy (at $T > 0$), because
of size quantization of the antiferromagnetic magnons inside the
islands. Enhanced Nernst effect can then occur. (A spin bag
should also be associated with a local depletion of magnetic
entropy. Only it may not be energetically as favorable if the
bag size is comparable to the object shown in Fig.1, as I have
argued above.) I expect energy savings when DCHAPHIs are
formed from the original singly-charged objects, but the total
entropy should at the same time decrease. Thus these new objects
are expected to form only below some charateristic temperature
which I identify as the pseudogap onset temperature
$T^*$.~\cite{pseudogap} The observed pseudogap for $T<T^*$ is
identified as half of the formation energy of a DCHAPHI. I
expect this energy to not change much with $T$. The main effect
of rising $T$ is then to only reduce the number of such
objects. {\it This is consistent with tunneling data in the
pseudogap regime~\cite{pseudogap,tunneling} where one sees a
pseudogap which does not shrink in magnitude much with rising
$T$, but only gradually gets filled in}. (This fact is against
interpreting $T^*$ as the mean-field $T_c$, the $T_c$ for
incoherent pairs, or the highest of a distribution of $T_c$.)

At any $x > x_{c1}$ the singly-charged object is a linear
combination of a singly-charged magnetic polaron and a bare
charge carrier, because short-range AFO fluctuates in and out of
existence. But the DCHAPHIs can exist only when local short-range
AFO is present. Thus the so-identified $T^*$ is a monotonically
decreasing function of $x$, (as is observed,) because the
average formation energy of a DCHAPHI should be a decreasing
function of $x$, due to the fact that in any given region of
space, short-range AFO fluctuates into existence for smaller
fraction of the time when $x$ is larger. This view is consistent
with the fact that the observed pseudogap decreases with
$x$.~\cite{pseudogap}

Figure 1 gives a snapshot of such a DCHAPHI, but it should be
emphasized that the size and shape of a DCHAPHI are actually both
fluctuating. So Fig. 1 gives only its typical size and shape.
Later I will argue why the average size of such an object is
actually smaller than shown here.

\begin{figure}[th]
\centerline{\psfig{file=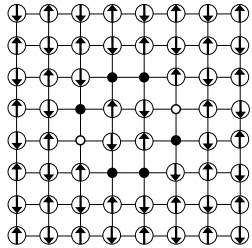,width=8cm}}
\vspace*{-4pt}
\caption{A snapshot of a doubly-charged anti-phase island in a
short-range antiferromagnetic background. An up (down) arrow
indicates a spin-up (down) electron at that site. Filled circle
indicates an electron at that site but $<s_z> = 0$. Open circle
indicates a hole, or no electron at that site.}
\end{figure}

If these DCHAPHIs are not trapped, they should be mobile objects
through deformations of their boundaries. HTSC in the cuprates
can then be identified as the Bose-Einstein condensation (BEC)
of these ``preformed pairs''. The observed continuity of the
tunneling gap across $T_c$ supports this view. One can also
qualitatively undertand why $T_c$ first rises from zero at
$x_{c2} \simeq 0.05$, reaches a peak at $x = x_{\rm opt}$ for
optimum doping, and then falls to zero at $x = x_{c3} \geq
0.25$.~\cite{Tc} The initial rise is because BEC temperature is
a rising function of boson concentration, which in our case is
the concentration of the DCHAPHIs, which at $T = 0$ clearly is
an increasing function of $x$ for small $x$. (See later
discussion for $x >$ about 0.1.) The downward bending of $T_c$
to reach a peak and the fall of $T_c$ after the peak can be
understood as due to the fall of $T^*$, which leads to reduced
number of ``preformed pairs'' at a given temperature.

\begin{figure}[th]
\centerline{\psfig{file=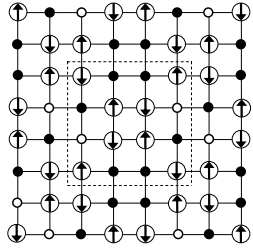,width=8cm}}
 \vspace*{-4pt}
\caption{Close packing of the DCHAPHIs shown in Fig.1. The dotted
square can be repeated to generate the whole structure. It
encloses sixteen sites and two holes. Hence this configuration
corresponds to $x = 0.125$.}
\end{figure}

A common difficulty in this type of theories, including this
theory, the spin-bag theory, and all preformed-pair theories, is
the size of the objects involved, since these objects generally
can not exist near and beyond close packing. Fig. 2 shows a close
packing density for the DCHAPHIs, and it only corresponds to $x
= 0.125$. This is far below the observed $x_{c3} \simeq 0.25$.
Even at this density the DCHAPHIs are probably no longer
energetically favored to form, since there is little AFO
outside their boundaries. We conclude that at $T = 0$ and for
$x$ beyond some $x$ value around 0.1, the number of DCHAPHIs can
not continue to increase with $x$. The remaining charge carriers,
--- there are more of them for larger $x$, --- must exist in
another form, such as the singly-charged objects mentioned
earlier. These singly-charged objects can weaken short-range AFO,
contributing to the lowering of the DCHAPHI formation energy,
$T^*$, $T_{\nu}$, and $T_c$, more for larger $x$ in
this region. (Note that at $T$ substantially above zero, there
are less DCHAPHIs, so the close-packing limitation is less
stringent. Thus for a limited range of $x$ beyond about 0.1, the
number of DCHAPHIs can still increase with $x$. This might
account for the continued rise of $T_c$ with $x$ until the
latter reaches $x_{\rm opt} \simeq 0.15$. The fact that even at
$T=0$ not all doped charges form DCHAPHIs is not as serious as
it might appear to be, since already in the BCS theory, only a
small fraction of the electrons near the Fermi surface form
Cooper pairs. Yet it does not prevent a BCS superconductor from
exhibiting strong superconducting properties such as perfect or
nearly perfect diamagnetism.)

\begin{figure}[th]
\centerline{\psfig{file=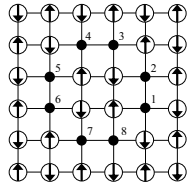,width=8cm}}
\vspace*{-4pt}
\caption{The eight boundary sites, numbered 1 through 8,
of the anti-phase island shown in Fig.1.}
\end{figure}

To understand the experiment of Iguchi et al.,~\cite{Iguchi-etal}
we first show that these DCHAPHIs are diamagnetic. Figure 3 shows
the eight boundary sites of an anti-phase island shown in Fig. 1,
labled 1 throught 8, on which a doped charge carrier reside.
[Actually there are two such charges on them, but we need only
consider one here.] The hopping matrix elements between sites 1
and 2, 3 and 4, 5 and 6, and 7 and 8 are just the hopping matrix
element $-t$ (with $t>0$) in the usual $t-J$ model, which has
been shown to be equivalent to the Hubbard model for large
$U$.~\cite{anderson} However, the effective hopping matrix
elements between sites 2 and 3, 4 and 5, 6 and 7, and 8 and 1,
denoted as $-t'$, should satisfy $t'<0$, because they are the
result of third-order processes with one illustrated in Fig. 4,
(and another one which is just this one in a different order,)
involving two negative energy denominators and three matrix
elements in the numerator, two of which are $-t<0$, and one of
which is $J>0$.

\begin{figure}[th]
\centerline{\psfig{file=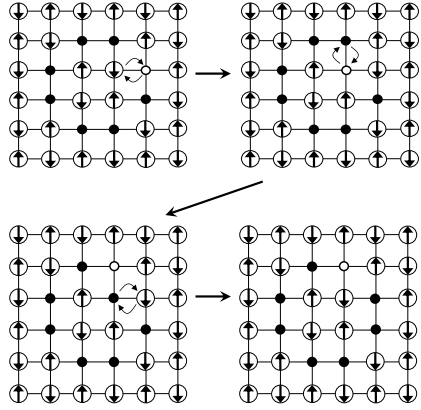,width=8cm}}
 \vspace*{-4pt}
\caption{A third order process which allows a doped charge
to hope from site 2 to site 3 defined in Fig. 3.}
\end{figure}

The effective Hamiltonian is therefore
\begin{eqnarray}
{\cal H} &=& -t(\hat c^{\dagger}_2\hat c_1 + \hat
c^{\dagger}_4\hat c_3 + \hat c^{\dagger}_6\hat c_5
+ \hat c^{\dagger}_8\hat c_7)\nonumber\\
&+& |t'|(\hat c^{\dagger}_3\hat c_2
+ \hat c^{\dagger}_5\hat c_4 + \hat c^{\dagger}_7\hat c_6 + \hat
c^{\dagger}_1\hat c_8) + h.c.\,.
\end{eqnarray}
The {\it non-degenerate} ground state of this Hamiltonian is
$(1, 1, -1, -1, 1,$ \noindent $1, -1, -1)^{\dagger}$, i.e., a
$d$-wave state, with energy $-(t + |t'|)$. It is filled by two
holes (or electrons) forming a singlet state, so there is
clearly no spin contribution to the magnetic susceptibility.
This eigen-vector is real, so its angular momentum expectation
value is zero. The state is then diamagnetic. (This is basically
Larmar diamagnetism, also known as Langevin
susceptibility.~\cite{AshcrftMermn}) Extending the above 8-site
model to include the effect of an external magnetic field has
confirmed this conclusion. The parameter $t'$ should be left as
an independent parameter of the theory, since any
next-nearest-neighbor hopping term in the original Hubbard model
can also contribute to $t'$ directly. These DCHAPHIs can probably
undergo phase separation, since the phase-separation argument of
Emery et al.~\cite{Emery_etal} should apply to these objects as
well as the charged objects they considered. (An alternative
possibility is a phase separation of the two types of charged
objects without creating an inhomogeneous charge distribution.)
The diamagnetic patches observed by Iguchi et al. are then
qualitatively explained. The increased total area of such patches
as $T$ is lowered from $T^*$ is then simply due to the formation
of more such diamagnetic objects. Figure 4 also shows that the
size of a DCHAPHI is smaller than is shown in Fig. 1 at any
transiant moment when a doped charge carrier is hopping between
such two sites as 2 and 3. (If a third-order process were
to involve a site outside the boundary, two neighboring spins
would become parallel. Such intermediate states would have higher
energy.)

Finally, let us look at the enhanced Nernst effect observed by
Xu et al.~\cite{Xu-etal} Vortices can give rise to such an
effect, since each vortex carries a magnetic flux along $B_z$,
and also carries excess entropy inside its core. These are the
two essential ingredients for generating an enhanced Nernst
effect. Another situation can also give rise to this effect with
the same sign: That is when there are localized diamagnetic
objects in the system with a depletion of entropy in the spatial
regions occupied by them. This is precisely the present
situation, since the DCHAPHIs must also be associated with a
depletion of entropy inside the spatial regions occupied by them
because of size quantization of the antiferromagnetic magnons
inside the islands. Xu et al. found that $T_{\nu}$ first rises
sharply at an $x$ value very close to $x_{c1}$, reaching
a peak at an $x$ larger than $x_{c2}$ but below $x_{\rm opt}$
(i.e., at $\sim$ 0.1 for LSCO), and then to fall gradually as $x$
is increased further. The gradual fall is simply due the gradual
fall of $T^*$ in this $x$ range. In this range $T_{\nu}$ is
below $T^*$ probably because a small $x$-dependent number of
these DCHAPHIs are trapped by inhomogeneous potential to become
immobile, and therefore can not contribute to the Nernst effect.
Another possible contributing factor is that the two types
measurements may have different sensitivity on a very low
concentration of such objects. The patching of a low
concentration of DCHAPHIs may also be at least partially
responsible for the difference. The initial sharp rise of
$T_{\nu}$ is most-likely also due to the potential traps: A
very low concentration of such objects can become all
trapped and immobile, and therefore can not contribute to the
Nernst effect, even though they can still give rise to a
pseudogap in tunneling with deeper center dip at lower $T$,
because their number still increases with lower $T$. The number
of such traps is a rapidly decreasing function of $x$
for reasons already given. Clearly $T_{\nu}$ has to increase
sharply so that the number of DCHAPHIs at $T > T_{\nu}$ can
still be so low that they can be all trapped. Since raising $T$
can further release some trapped objects, this sharp rise of
$T_{\nu}$ can only be even sharper. We thus have a qualitative
understanding of the initial rapid rise of $T_{\nu}$.

We can not make any quantitative predictions for testing purpose
at the present stage. Nevertheless, there can still exist
additional qualitative signatures of this theory besides the ones
already discussed. One important confirmation of this theory
would be the observation of singlet or triplet excited states of
the DCHAPHIs. But they may not exist as stable objects. Thus for
strong support of this theory one should look for patched
pseudogap using STM, in the same sample type, spatial and
temperature regions where patched diamagnetism is observed.
Another strong support would be if the noise spectrum in the
Nernst signal is found to be drastically different above and
below $T_c$. (The flux in a vortex is fixed, but the number of
vortices is proportional to $B_z$. The flux in a DCHAPHI is
proportional to $B_z$, but its number should be essentially
independent of $B_z$.)

{\it Note added after refereeing process:} We have completed a
mean-field study of the Hubbard model which confirms the
existence and stability of a DCHAPHI and the conversion of a
doubly-charged spin-bag to a DCHAPHI when the Coulomb repulsion
between the two charges are turned on. [Q. Wang and C.-R. Hu,
unpublished.]

%\begin{acknowledgments}
The author would like to acknowledge support from the Texas
Center for Superconductivity and Advanced Materials at the
University of Houston.
%\end{acknowledgments}

\end{document}